\documentclass[conference]{IEEEtran}
\IEEEoverridecommandlockouts
\usepackage{float}
\usepackage{comment}
\makeatletter
\let\newfloat\newfloat@ltx
\makeatother
\usepackage[utf8]{inputenc}
\usepackage{xcolor}
\usepackage{graphicx}
\usepackage{amsmath}
\usepackage{amsthm}
\usepackage[normalem]{ulem}
\usepackage{amssymb}
\usepackage{url}
\usepackage{soul}
\usepackage{subfig}
\newcommand\redout{\bgroup\markoverwith
{\textcolor{red}{\rule[.5ex]{2pt}{0.4pt}}}\ULon}
\usepackage[ruled,linesnumbered,vlined]{algorithm2e}
\usepackage{etoolbox}

\newcounter{algoline}
\newcommand\Numberline{\refstepcounter{algoline}\nlset{\thealgoline}}
\AtBeginEnvironment{algorithm}{\setcounter{algoline}{0}}
\usepackage{booktabs}
\usepackage{multirow}
\let\originalleft\left
\let\originalright\right
\renewcommand{\left}{\mathopen{}\mathclose\bgroup\originalleft}
\renewcommand{\right}{\aftergroup\egroup\originalright}

\newcommand{\Expt}[1]{\mathbb{E}\left[#1\right]\xspace}

\newcommand{\para}[1]{\vspace{0.1em}\noindent\textbf{#1.}~}

\def\BibTeX{{\rm B\kern-.05em{\sc i\kern-.025em b}\kern-.08em
    T\kern-.1667em\lower.7ex\hbox{E}\kern-.125emX}}
\begin{document}

\title{Distort to Detect, not Affect: Detecting Stealthy Sensor Attacks with Micro-distortion%
}

\author{\IEEEauthorblockN{Suman Sourav\IEEEauthorrefmark{1}, Binbin Chen\IEEEauthorrefmark{2}\IEEEauthorrefmark{1}}
\IEEEauthorblockA{\IEEEauthorrefmark{1}Advanced Digital Sciences Center, Singapore\\ \IEEEauthorrefmark{2}Singapore University of Technology and Design, Singapore} 
Email: suman.s@adsc-create.edu.sg, binbin\_chen@sutd.edu.sg}

\maketitle
\IEEEpubidadjcol

\begin{abstract}
In this paper, we propose an effective and easily deployable approach to detect the presence of stealthy sensor attacks in industrial control systems, where (legacy) control devices critically rely on accurate (and usually non-encrypted) sensor readings. Specifically, we focus on stealthy attacks that crash a sensor and then immediately impersonate that sensor by sending out fake readings. We consider attackers who aim to stay hidden in the system for a  prolonged period. To detect such attacks, our approach relies on continuous injection of ``micro distortion'' to the original sensor's readings. In particular, the injected distortion should be kept strictly within a small magnitude (e.g., $0.5\%$ of the possible operating value range), to ensure it does not affect the normal functioning of the ICS.  Our approach uses a pre-shared secret sequence between a sensor and the defender to generate the micro-distortions.  One key challenge is that the micro-distortions injected are often much lower than the sensor's actual readings, hence can be easily overwhelmed by the latter. 
To overcome this, we leverage the observation that sensor readings in many ICS (and power grid in particular) often change gradually in a significant fraction of time (i.e., with small difference between consecutive time slots). We devise a simple yet effective algorithm that can detect stealthy attackers in a highly accurate and fast (i.e., using less than 100 samples)  manner. We demonstrate the effectiveness of our defense using real-world sensor reading traces from two different smart grid systems.  
\end{abstract}

\section{Introduction}

With the wide adoption of Industrial Control Systems (ICS) in different critical infrastructures like smart grids \cite{mcdaniel2009security}%
, water / waste-water treatment systems \cite{weiss2014industrial}%
, and nuclear power plants \cite{cho2015cyberphysical}, the issue of ICS security has become increasingly important. 
High-profile attacks such as Stuxnet~\cite{nourian2015systems} and the Ukraine power grid blackouts~\cite{lee2017crashoverride} have shown that adversaries can wreak havoc by compromising selected ICS devices. In particular, the attackers in those incidents were able to stealthily compromise critical devices and gathered key information about the system over a period of time. The well-prepared attackers then launched their attacks to cause devastating physical impact on the critical infrastructure. 

Here, we consider such stealthy attackers who have already gained some foothold in the system and aim to stay hidden for a prolonged period (e.g., to cause long-term damage or to wait for the planned date to launch attacks). 
As part of the attack execution, we assume such attackers have crashed some critical sensors in the ICS. We assume these sensors communicate with the other devices in the ICS through a network. %
Attacker(s) can remotely crash such sensors by exploiting some faults in their firmware --- such faults are more common than faults that allow an attacker to take over the full control of a sensor.
Once an original (i.e., real) sensor is crashed, the attacker will immediately impersonate that sensor by sending fake sensor readings to the network. Since most sensor transmissions in today's ICS are not protected by encryption (nor by any authentication schemes), the other devices that depend on these sensor readings cannot distinguish between the real sensor and the impersonated sensor. The attackers want to remain stealthy, so if there are any intrusion detection mechanisms in the ICS, the attacker will use their best knowledge about both the system's operation behavior and the intrusion detection rules to carefully craft their fake sensor readings, making them look normal and indistinguishable from the real sensor's readings.  

While there have been significant advances in securing ICS against attacks to the sensor readings (e.g.,~\cite{6545301, 5394956, 7011170, 7039974, 10.1145/3274694.3274748}),  %
many of the proposed solutions require major upgrading of the existing ICS, e.g., by introducing authentication or/and encryption schemes at both the sensors and the controllers, or work based on assumptions that may not hold for advanced and persistent attackers, e.g., assuming the attackers do not know about the system's operational behavior or cannot observe some unique features of the sensors before launching the attack.

In this work, we seek to design a practical solution to secure legacy ICS against stealthy sensor attacks. The solution should be based on assumptions that even advanced attackers cannot easily bypass, %
while making minimal changes to a legacy ICS. Furthermore, 
the solution should have a negligible impact on the functioning of all legacy devices in the ICS.
{\noindent \bf Injecting micro-distortion based on a sequence of secret:}
The key idea of our approach is to 
continuously introduce a very small distortion (which we will call ``micro-distortion'' hereafter) to the readings of the sensor that we want to protect. 
To avoid affecting the normal functioning of the ICS, the injected distortion should be kept within a very low magnitude $\epsilon$ (e.g., 0.5\% of the possible operational value range). To use the presence of such distortion to authenticate the sensor, we generate the distortion based on a secret that is shared only between the sensor and a defender. The secret contains a sequence of binary values 0 and 1 (i.e., a one-time pad), one value to be used for each time instance.
Given $k_i$ (the secret key for a time slot $i$),
$d_i$ (the actual reading of sensor at time $i$) is changed to 
$d'_i = d_i + (2k_i-1) \epsilon$. Specifically, an increment or decrement by $\epsilon$ based on the value of $k_i$.

A potential way to introduce micro-distortion to the reading of a sensor in an ICS is to do it \emph{physically}. Say, by deploying a micro-actuator that will physically introduce the micro changes to the underlying system, so that the original sensor will pick that up in its reading. For example, consider an electrical meter that measures the power consumption of a system with 50 kilowatt of peak load. We can introduce a securely controllable load that can dynamically vary its load by $\pm 250$Watt (or $\pm0.5\%$ of the peak load) based on the secret. For the purpose of this study, digital manipulations at the sensor itself is not allowed, as this would require an upgradation of all the sensors, increasing the cost of implementation.

{\noindent \bf Detecting attack based on presence of micro-distortion:}
The secret sequence kept between a sensor and a defender
forms the basis for the defender to distinguish between the real sensor and a fake one. With the secret, we aim to introduce a statistical difference via the micro distortion. However, as we will show later,
one key challenge  
for detection using micro-distortion is that, by design, the magnitude of the micro-distortion is much lower than the sensor's actual readings, hence can be easily overwhelmed by the latter. 
For example, if the actual readings are drawn uniformly randomly and independently from all the possible range of values, and when the micro-distortion $\epsilon$ equals to $0.5\%$ of the possible range, it 
will require more than 80,000 samples in order to reduce both the false positive (FP) and false negative (FN) rate below $1\%$. 
Even if the sensor reading is sent every second, this translates to almost one whole day's delay for detecting an attack.
If the sensor reading is sent only every minute, it further inflates the required detection delay to around $2$ months.
To overcome this, we leverage the observation that sensor readings in many ICS (and power grid included) often change gradually in a significant fraction of time (i.e., consecutive measurements have small difference).
Based on this observation, we devise an effective {\it Filtered-$\Delta$-Mean-Difference} algorithm that is based on statistical gauges calculated over the consecutive change of the sensor readings, instead of the raw sensor reading sequence directly. 
We demonstrate the effectiveness of our defense using real-world sensor reading traces from two different smart grid systems --- one monitors the power generation from a solar farm, while the other monitors the household power usage. Our experiments confirm that our algorithm can detect stealthy attackers in a highly accurate (with false positive and false negative rate at $1\%$ or even lower) and fast (i.e., using less than 100 samples)  manner, and can achieve more than $100$ times deduction in terms of the detection delay compared to the baseline detection approach.

In contrast to many existing approaches, our solution is extremely easy to deploy, light-weight and low cost,  specifically for resource constrained legacy ICS systems.  Once a physical secret key has been shared with each sensor, any defender with some reasonable computing power, would be able to detect the presence of stealthy attackers (if any). Also, note that, the secret key only needs to be shared between the sensor and the defender.  Other components (e.g., controllers) that use the sensor's data use it directly without requiring to filter out the injected distortions, and as such do not need to know the sensor's secret key.  This not only reduces the chance of leakage, but makes deployability easier as none of these components would require any upgrading.  %

The key contributions of this work include:
\begin{itemize}
\item We propose a simple yet effective micro-distortion based solution for detecting stealthy sensor attacks. Our solution can be easily implemented in legacy ICSes, with minimal change to the overall working of the system, while significantly improving security.
\item We propose a novel detection algorithm based on the filtered delta sequence, that considers the value difference between consecutive distorted sensor reading sequence.
Our algorithm leverages the observation that in many real ICS, sensor readings tend to change gradually in a significant fraction of time. This allows our algorithm to out-perform %
a baseline algorithm that directly uses the original sensor reading for attack detection by a few orders of magnitude. 
\item We give detailed experimental case studies based on real-world traces from two different power systems, demonstrating the effectiveness of our detection algorithm.
\end{itemize}

\section{Related Work}

Attack detection in ICS, as opposed to traditional fault detection, is more challenging as the adversaries can be persistent, intelligent, and stealthy. They can gather and use the knowledge of the system to remain undetected. %
While attacks that manipulate the controller logic in ICS can be detected using software attestation~\cite{chen2017secure} or deception technology~\cite{mashima2017towards}, existing solutions to detect sensor reading manipulation often face challenges when dealing with stealthy attackers.
For example, traditional bad-data detection techniques,  such as the largest residue test \cite{abur2004power}, may fail to detect an intelligent attacker who changes state estimation of the system while introducing errors within the range space of the observation matrix. Similarly, approaches such as \cite{biswas2019electricity} cannot work when redundant sensors that measure the same physical phenomenon are all compromised. %

%

%

\begin{comment}
From the perspective of control theory, numerous detection techniques have been proposed  %
for detecting intelligent attacks like false data injection attack and replay attack \cite{LI2019474,7322210}.
\end{comment}
%
%
Several watermarking-based authentication mechanisms where an actuator superimposes a random signal, known as the watermark, on the control policy-specified input while checking for an appropriate response from the sensors, were studied in \cite {5394956, 7011170}. %
There, the physical watermarking is added to the control output and studied specifically in the context of replay attacks, where an attacker just replays previously observed measurements of a system.  In \cite{7039974}, the watermarking scheme is extended for false data injection attacks where adversaries have the power to substitute real measurements with generated stealthy signals and was further improved in \cite{7738534,  8287297}.  %
Though similar in concept to a shared key, most of these solutions are focused on designing watermarked control inputs to detect counterfeit sensor outputs.  Moreover, these solutions are specific to linear dynamical systems described by time-invariant parameters and rely on accurate modelling of the system states.  
In contrast, our work pertains to detecting attackers by adding micro-distortions directly at the sensors which can then be leveraged to detect stealthy attackers.
In \cite{10.1145/3274694.3274748}, the authors propose to authenticate sensors and detect data integrity attacks in CPSs by using a sensor's hardware characteristics along with physics of a process to create unique fingerprints for each sensor. {However, a persistent attacker who is patient enough to observe the fingerprints of the sensor before crashing it, could learn about the fingerprint and potentially use the device under its control to reproduce such learnt fingerprint when impersonating the crashed sensor. Also, as shown in \cite{10.1145/3274694.3274748}, 
the false positive (FP) and false negative (FN) rate this approach can achieve is around $5\%$, which may not be acceptable for settings where any false positive or negative incurs a high cost to deal with.}

Different approaches based on cryptographic primitives have also been proposed to address this problem. For example,  homomorphic encryption based solutions were proposed in  \cite{kim2016encrypting,  7403296,  min2019privacy} and public-encryption systems in  \cite{6485982, 6857845}. %
They utilize computationally heavy encryption/decryption algorithms which not only increases delay but also requires high upgradation cost for legacy systems. Other solutions, like \cite{8340689,  SEDA} rely on the installation of additional specialized equipment which can increase the upgradation costs significantly.  
\addtolength{\topmargin}{0.05in}
\section{Threat Model}

\begin{figure}[t!]
  \centering
    \includegraphics[width=0.95\linewidth]{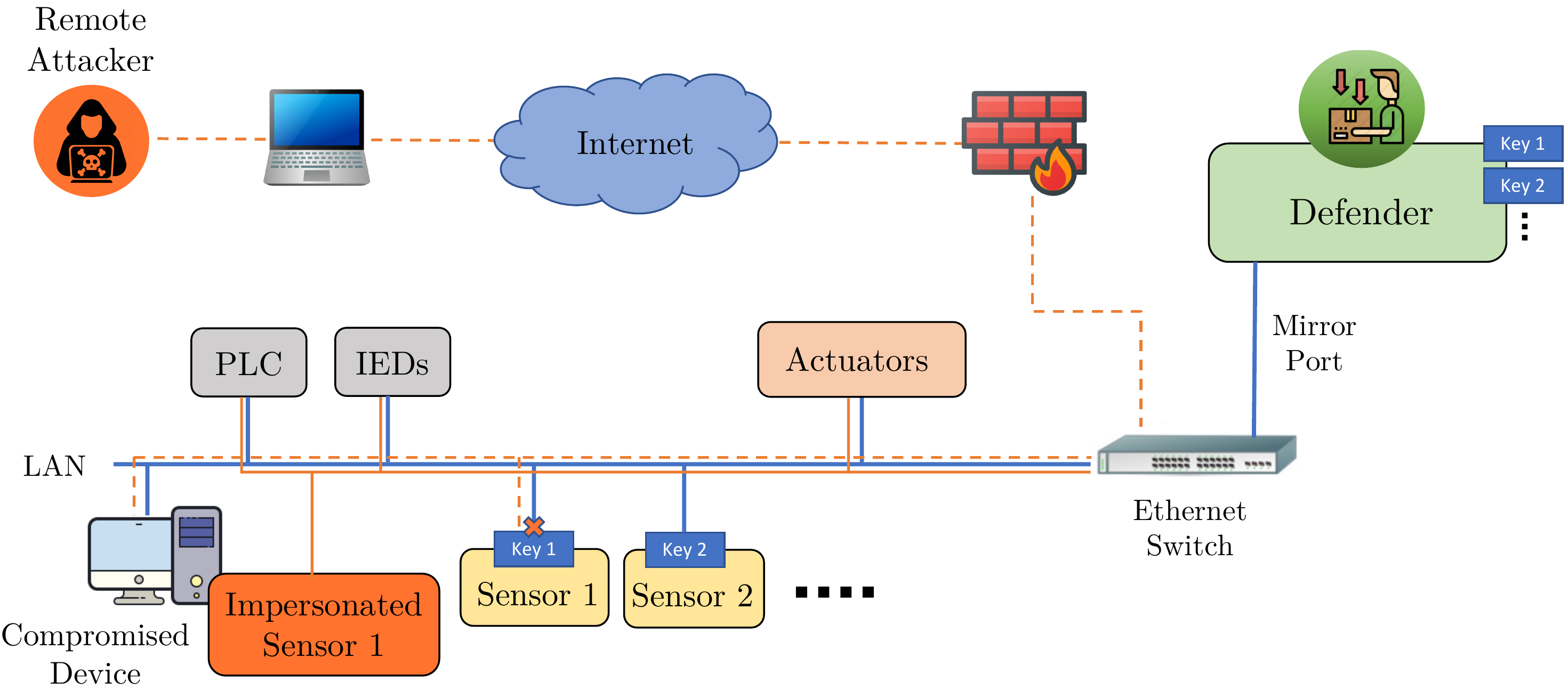}
  \caption{\small Our threat model over a typical ICS. The blue line represents the connection between the devices, the dashed red line represents the attack path where the remote attacker crashes Sensor 1 and immediately impersonates it via a compromised device. The red line shows the traffic from the impersonated sensor to the other connected devices through the Ethernet switch. In the absence of an attack, similar traffic flow exists between Sensor 1 and the other devices.\vspace{-3mm}}
  \label{fig:system}
\end{figure}

As shown in Figure~\ref{fig:system}, we assume that the attacker has gained a foothold (e.g., programmable logic controller, human-machine-interface, historian) in the ICS network, enabling him or her to collect physical knowledge about the system without being detected. 
We also assume the attacker can only compromise few sensors and is unable to compromise a large majority of the sensors.  %
This is a reasonable assumption because there are often different types of sensors in the system and they can be in different network zones. %
Also, larger the number of compromised devices, higher the risk of being detected. This goes against the attacker's aim to remain stealthy.

After taking a subset of the legitimate sensors offline, the attacker immediately impersonates those sensors. %
Any significant gap between the crashing step and the impersonating step may look suspicious and enable an IDS to detect the attack. Also, %
the attacker needs to keep injecting sensing readings on behalf of those crashed sensors on a regular basis.   

To make the attacker as strong as possible, we assume the attacker can observe any sensor for a significantly long duration, before taking control of the sensor. As such the attacker is assumed to have all historical data of the system, from which it can gain complete knowledge about the physical system. 
However, the attacker is unaware of the shared secret between a sensor and the defender.

Additionally, for the scope of this work, while staying stealthy,  an adversary can only modify sensor output of the compromised sensor and does not change control signals. 
    
\section{Detection Using Micro-distortion} \label{Sec:Detection_Algo}

Since each sensor is associated with some secret key,
one straightforward solution would be to use that key to encrypt the sensor's messages, or to authenticate either the sensor or its messages.
This, however, also requires the secret key to be shared with all the devices in the ICS that rely on the sensor data.
Overall, %
this requires all affected devices in the ICS to be upgraded to accommodate the introduced changes which might be rather costly.
Also, if the sensor's reading is consumed by multiple devices in a broadcast or multicast group, sharing a single key exposes it to a large attack surface. Any compromised member in the group can impersonate that sensor. Using asymmetric keys can mitigate this risk, but incurs much higher computation overhead on both the sensor and the other devices.

\subsection{A Strawman Design: Simple Mean Difference}
If one considers a perfectly stable noiseless system where the sensor readings remain constant for the detection period, %
it is easy to see that our approach can detect a compromised sensor extremely fast, by simply letting the defender checking for the pattern of the distortion based on the same secret.
An attacker without the knowledge of the secret would not be able to replicate the pattern, hence cannot bypass the detection. In fact, if the attacker just makes a random guess, in each slot, it has $50\%$ chance of guessing wrongly and therefore being detected. Hence, the probability that the attacker can remain undetected after $10$ slots is as low as $0.5^{10} < 0.1\%$. However,  under normal functioning of an ICS, the sensor readings are subject to the state changes in the ICS along with possible noise. With the possible sensor readings spanning a much wider (e.g., 200x) range of a micro-distortion, the micro-distortion becomes a negligible signal that easily gets overwhelmed by the magnitude of the actual sensor readings.

\begin{figure}[t!]
  \centering
    \includegraphics[scale=0.28]{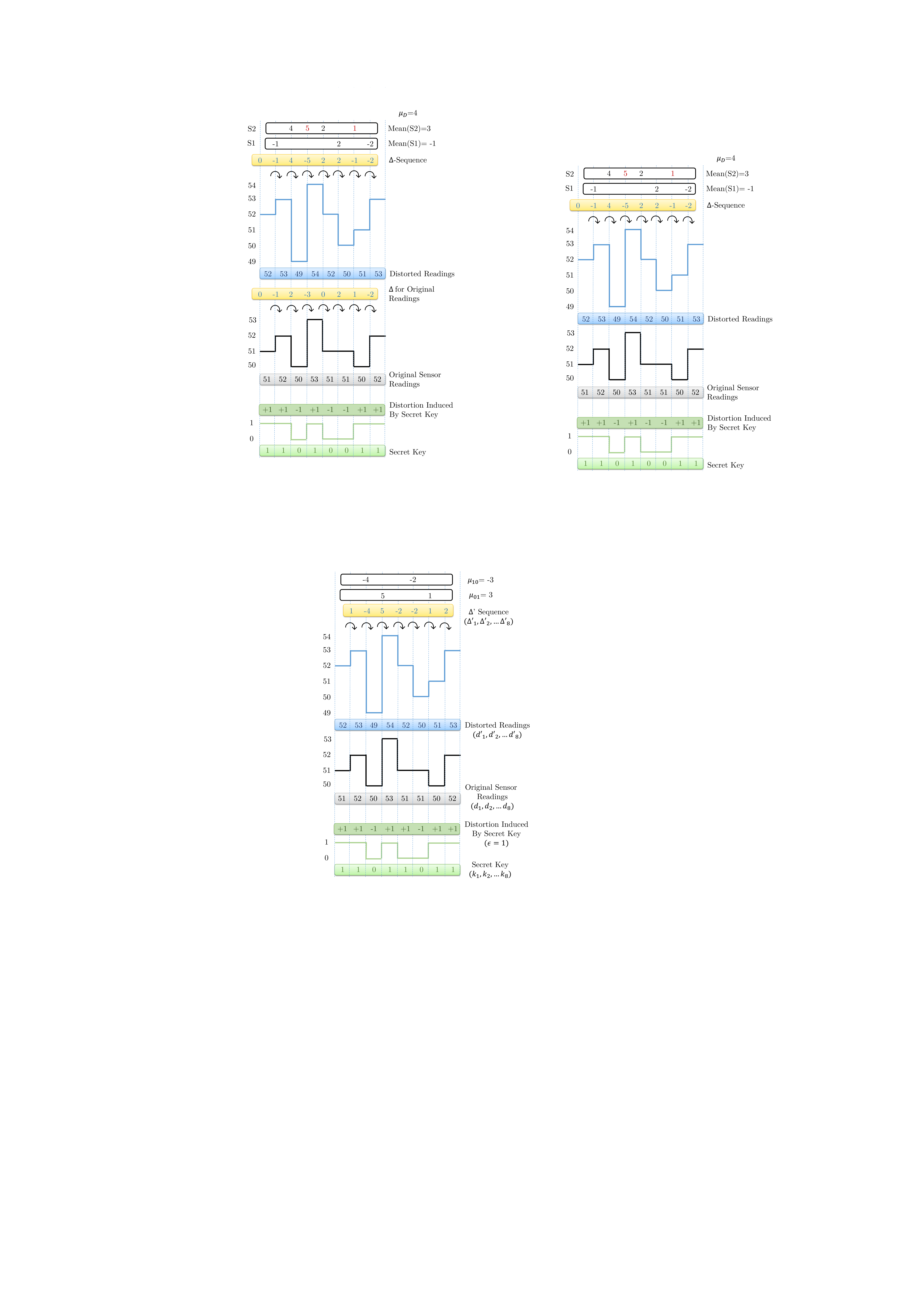}
  \caption{An Example illustrating the different notations used.}
  \vspace{-4mm}
  \label{fig:det_algo}
\end{figure}

\para{Simple Mean Difference} 
By the ``law of large numbers" \cite{dekking2005modern},
the mean value of a large number of observations made for a random variable 
approaches the random variable's expected value as more observations are taken. 
Consider the set of readings distorted as $d'_i=d_i+\epsilon$ as set $S_1$ (i.e., with corresponding secret key $k_i=1$), and the remaining set as $S_0$, i.e., with $k_i=0$ and corresponding reading distorted as $d'_i=d_i-\epsilon$). Since we select the two sets from all the readings over a time window based on the random secret, we can view the original readings $d_i$ from both sets as observations drawn from the same distribution. Hence, with a larger number of observations, the difference between the mean value of all distorted readings from set $S_1$ and the mean value of all distorted readings from set $S_0$ should approach $2\epsilon$. If the attacker does not know the secret key, it cannot introduce any statistical difference between these two sets. As such, the difference of the mean value for $S_0$ and $S_1$ should approach $0$ when under attack. 

This shows that the detection can eventually be achieved if the detector can examine a sufficiently large number of samples. The issue of simply relying on the law of large numbers, however, is that the detection can be rather slow. For example, if the actual readings are drawn uniformly randomly and independently from all the possible range of values, and when the micro-distortion $\epsilon$ equals to $0.5\%$ of the possible range, our evaluation result shows that it requires more than 80,000 samples in order to reduce both the false positive and false negative rate below $1\%$. Similarly, it needs more than 140,000 samples in order to further reduce that to below $0.1\%$. Even if the sensor reading is sent every second, this translates to almost one whole day for achieving $1\%$ false positive/negative rate, and nearly $40$ hours of readings for $0.1\%$ false positive/negative rate. If the sensor reading is sent only every minute, it will further inflate the required detection delay to around $2$ and $3.5$ months respectively.

\subsection{Our Design: Filtered-$\Delta$-Mean-Difference}

Fortunately, in real-world ICS, the sensor readings present some good statistical properties that allow much faster detection of such attacks. Specifically, in many ICS (and power grid in particular), physical properties being sensed (e.g., the amount of power being generated or consumed) tend to change in a gradual manner (i.e., with small differences between consecutive time slots) in a significant fraction of time.
Based on that, we propose a %
detection algorithm, where 
instead of comparing the difference of the mean of the sensor readings from sets $S_1$ and $S_0$ (as defined earlier), we look at the change in sensor reading between consecutive time slots, which we refer to as $\Delta$.

\para {Adding Micro Distortions} %
Each sensor determines the micro-distortion $\epsilon$ value based on the magnitude of the sensor readings (e.g., $< 0.5\%$ of the sensor's operating range).  For each sensor reading $d_i$,  if the corresponding key-value $k_i$ is $1$, then the sensor's reading is distorted by adding a value of $\epsilon$ to the reading; else is distorted by subtracting a value of $\epsilon$ from the reading.  Note that, adding distortions in this form maintains a zero mean distortion in the long run.

\para {$\Delta$-sequence creation} Given the distorted sensor reading sequence $d'_1, ....,  d'_n$ and the secret key sequence $k_1, ..., k_n$, 
we define the $\Delta$ sequence as $\Delta_1, ..., \Delta_{n-1}$ and $\Delta'$ sequence as $\Delta'_1, ..., \Delta'_{n-1}$, where
\[\Delta_i = d_{i+1} - d_i \text{ and } \Delta'_i = d'_{i+1} - d'_i\]
While $\Delta_i$ gives the difference between the original sensor readings in consecutive time slots, $\Delta'_i$  gives the difference between the distorted sensor readings in consecutive time slots.

\para{Data partitioning step} %
We define set $S_{01}$ as the collection of all moments $i$ such that $k_i=0$ and $k_{i+1}=1$ and we define set $S_{10}$ as the collection of all moments $i$ such that $k_i=1$ and $k_{i+1}=0$. We define set $S_{00}$ and set $S_{11}$ similarly.
It could be seen that for an $i$ that belongs to different sets, the relationship between the corresponding $\Delta_i$ and $\Delta'_i$ is different. Specifically, for $i\in S_{00}$ or $i\in S_{11}$, since the same distortion is applied to both $d'_i$ and $d'_{i+1}$, we can see that $\Delta'_i = \Delta_i$. On the other hand, $\Delta'_i = \Delta_i + 2\epsilon$ for $i\in S_{01}$, while $\Delta'_i = \Delta_i - 2\epsilon$ for $i\in S_{10}$.

Since each of the random key $k_i$ is drawn with equal probability from $0$ and $1$ in an independent manner, it is easy to see that a moment $i$ (in regard to $\Delta$ and $\Delta'$ sequence) has equal probability to be falling into one of the four sets $S_{01}$, $S_{10}$, $S_{00}$, and $S_{10}$. As the value in the $\Delta$ sequence does not depend on the value of the secret key sequence, we have:
\begin{eqnarray}
\Expt{avg(\Delta_i | i \in S_{01})}=\Expt{avg(\Delta_i | i \in S_{10})}\nonumber\\
=\Expt{avg(\Delta_i | i \in S_{00})}=\Expt{avg(\Delta_i | i \in S_{11})}\nonumber
\end{eqnarray}
$$\text{Consider the gauge }x=avg(\Delta'_i | i \in S_{01}) - avg(\Delta'_i | i \in S_{10}) $$
We have:
\begin{eqnarray}
& & \Expt{x}\nonumber = \Expt{avg(\Delta'_i | i \in S_{01}) - avg(\Delta'_i | i \in S_{10})} \nonumber\\
& = & \Expt{avg((\Delta_i + 2\epsilon) | i \in S_{01}) - avg((\Delta_i - 2\epsilon) | i \in S_{10}})  \nonumber\\
& = & 4 \epsilon + 
\Expt{avg(\Delta_i | i \in S_{01})} - \Expt{avg(\Delta_i | i \in S_{10})}  \nonumber \\ & = &4 \epsilon \nonumber
\end{eqnarray}

\para{Detection Condition} As shown, if we calculate the difference between the mean of all $\Delta'_i$ in set $S_{01}$ and that in set $S_{10}$ as $x$, the expected value of $x$ should be $4\epsilon$ when there is no attack. In comparison, in case of an attack, as we assume that the attacker does not know the value of $k_i$, it cannot introduce any significant statistical difference between the set $S_{01}$ and $S_{10}$. As a result, in this case, the expected value of $x$ should approach $0$. In other words, we can use the expected value of $x$ to differentiate between the attack and non-attack cases.

In particular, we calculate the $\Delta$-mean-difference $\mu_{01} = avg(\Delta'_i | i \in S_{01})$ and $\mu_{10} = avg(\Delta'_i | i \in S_{10})$. Thereafter, we check whether {$\mu_{01} - \mu_{10} \in \{2\epsilon, 6\epsilon\}$} (as the difference concentrates near the expected value of $4\epsilon$) and raise alarm to detect attack if the condition is not satisfied.

While this seems similar to using the expected difference of the mean value between $S_0$ and $S_1$ (the procedure, we refer to as `Simple Mean Difference'), the benefit of calculating using $\Delta'$ sequence is that, for many ICS, the absolute value of the elements in the $\Delta'$ sequence can be significantly lower than those in the distorted reading sequence (i.e., the $d'$ sequence). This is because in many ICS (including many power grid systems), while a particular physical measurement can have readings that span a large range (e.g., the peak power generation or consumption in an energy system can be $10\times$ or even $100\times$ bigger than its non-peak period), it turns to change gradually at most times. Thereby, making the distribution of value in $\Delta$ and $\Delta'$ sequence concentrate much heavily towards smaller values, i.e., the variance of the corresponding sequence is much smaller. The smaller values, in turn, make it possible to use a small number of samples to approach a given (small) neighborhood of the expected value with a higher probability.

\para{Filtration Step} Though the above steps, provide a complete detection algorithm, in practice, filtering out some high $\Delta'_i$ values %
(considering absolute values of $\Delta'$) that can cause high variance in the $\Delta$-sequence can often result in significant improvements. %
For systems with intermittent large abrupt changes, %
even though this filtration would reduce the sample size, it would significantly bring down the variation as well, making it much easier for attack detection while also improving the accuracy. Consequently, the $|\Delta'_i|$ readings that are greater than a particular threshold $\Delta_{th}$ are removed from consideration, where $||$ represents the absolute value function. %
$\Delta_{th}$ is based on the past (correct) operation of the sensor and is determined in %
a way that the number of $\Delta_i$ readings removed is not too much for the time duration considered, i.e., $n$.

However, an attacker might take advantage of such a filtration procedure by introducing high noise to faked sensor outputs which would likely result in most of the noisy data being filtered out, thus delaying the detection of the attacker. Such attackers can be checked by choosing another threshold $m$ (based on the $\Delta_{th}$ and system under consideration) which ensures that there is always sufficient $\Delta'_i$'s that get through even after the filtration step when there is no attack. %

See Algorithm \ref{alg:detection} for pseudocode for the `filtered-$\Delta$-mean-difference algorithm'. In the absence of the filtration step, we refer to the algorithm as `$\Delta$-mean-difference algorithm'.

\SetNlSty{textbf}{}{:}%
\IncMargin{.2em}%
\setlength{\textfloatsep}{0pt}
\begingroup
\LinesNumberedHidden
\begin{algorithm}[!t]
\DontPrintSemicolon
\caption{Filtered-$\Delta$-Mean-Difference Algorithm.}
\label{alg:detection}
 \KwIn{Given a time, denote the latest $n$ time slots as time slot $1$ to $n$, with corresponding pre-shared secret $\{k_1, k_2, \dots, k_n\}$ and the sensor readings $\{d'_1, d'_2, \dots, d'_n\}$ (which is supposed to be distorted according to the secret when there is no attack.}
 \KwOut{Raise an alarm if detecting the presence of an attacker in the given time interval.}
\Numberline \textbf{$\Delta'$-sequence creation: } 
Create $\Delta'$-sequence such that 
  $\Delta'_i = d'_{i+1} - d'_{i},  \forall i=1, \ldots, n-1$\\
\Numberline\textbf{Filtration Step: } All rows for which $|\Delta'_i| > \Delta_{th}$ are removed from consideration. \\
\Numberline\textbf{Data Partitioning Step: } Identify set $S_{01}$ as all time indices $i$ such that $k_i=0$ and $k_{i+1}=1$. Identify set $S_{01}$ as all time indices $i$ such that $k_i=1$ and $k_{i+1}=0$.\\ 
\Numberline \If{size of remaining $\Delta'_i$ with $i \in S_{01} \cup S_{10}$ is less than a threshold $m$}
								{Raise alarm to detect attack.}
\Numberline \Else{Calculate $\mu_{01} = avg(\Delta'_i | i \in S_{01})$ and $\mu_{10} = avg(\Delta'_i | i \in S_{10})$. \\
\If{$\mu_{01} - \mu_{10} \notin \{2\epsilon, 6\epsilon\}$}{Raise alarm to detect attack.} }
\end{algorithm}
\endgroup

\begin{comment}
\begingroup
\LinesNumberedHidden
\begin{algorithm}[!ht]
\DontPrintSemicolon
\caption{Detection Algorithm.}
\label{alg:detection}
 \KwIn{Pre-shared key of the form $\{b_1, b_2, \dots, b_k\}$ and a batch of $k$ consecutive sensor readings $\{d_1, d_2, \dots, d_k\}$ .}
 \KwOut{A boolean variable determining the presence/ absence of an attacker for the given block.}
%
\Numberline \textbf{$\Delta$-sequence creation: } 
Create $\Delta$-sequence such that 
  $\Delta_i = d_{i-1} - d_{i},  \forall i \neq 0$; $\Delta_0 = 0$\\
\Numberline\textbf{Filtration Step: } All rows for which $|\Delta_i| > \Delta_{th}$ are removed from consideration. \\
\Numberline\textbf{Data Partitioning Step: } Create set $S1$ containing all $\Delta_i$ such that either both $b_{i-1}$ and $b_i$ equals $0$, or both equals $1$.
To set $S2$ add all $\Delta_i$ such that $b_{i-1} = 1$ and $b_i = 0$; also, add $(-1 * \Delta_i)$ for all instances where  $b_{i-1} = 0$ and $b_i = 1$. \\
\Numberline \If{size of $S1 < k'$ or size of  $S2 < k'$}
								{Raise alarm to detect attack and exit.}
\Numberline Let $\mu_{S1}$ and $\mu_{S2}$ be the mean of all the values in sets $S1$ and $S2$ respectively.  \\
\Numberline Calculate Mean Difference $\mu_D = \mu_{S2} - \mu_{S1}$. \\
\Numberline\If{$\mu_D \in \{\epsilon, 3\epsilon\}$}{
                Continue to new input batch.
    }
\Numberline\Else {Raise alarm to detect attack and exit.}
\end{algorithm}
\endgroup

\end{comment}

\begin{table}[!h]
\small
\begin{center}
\begin{tabular}{|c|c|c|}
\hline
& Power (W) & $\Delta$-sequence (W)\\ \hline
Maximum Value & 17206.0  &  14351.0\\ \hline
Minimum Value &  225.0  &  0.0\\ \hline
Average Value & 1104.0733  &  13.611\\ \hline
Median Value &  775.0  &  3.0\\ \hline
\end{tabular}
\end{center}
\caption{Table showing the power usage data statistics for House 1 and its associated $\Delta$-sequence.}
\label{tab:house_data}
\end{table}

As by-product benefit of our solution, observe that now the one-time pad is not sent in its clear form through any communication between a sensor and the defender. There always exists the possibility of misinterpretation of a natural change of the sensor reading as a micro distortions to any wire-tapper or observer. So, even when the algorithm generating the one-time pad is weak, %
our use of the one-time pad hidden within the naturally occurring changes will make it harder for the attacker to exploit the weakness.

\begin{table*}[ht!]
\small

\begin{center}
\begin{tabular}{|c|c|c|c|c|c|c|c|c|c|}
\hline
\multirow{2}{*}{$n$} & \multicolumn{3}{c|}{\begin{tabular}[c]{@{}c@{}}Simple Mean Difference\end{tabular}} & \multicolumn{3}{c|}{\begin{tabular}[c]{@{}c@{}}$\Delta$-Mean-Difference\end{tabular}} & \multicolumn{3}{c|}{\begin{tabular}[c]{@{}c@{}}Filtered-$\Delta$-Mean-Difference\end{tabular}} \\ \cline{2-10} 
 & \begin{tabular}[c]{@{}c@{}}FP\%\end{tabular} & \begin{tabular}[c]{@{}c@{}}FN\%\\ (EDA)\end{tabular} & \begin{tabular}[c]{@{}c@{}}FN\%\\ (RDA)\end{tabular} & \begin{tabular}[c]{@{}c@{}}FP\% \end{tabular} & \begin{tabular}[c]{@{}c@{}}FN\%\\ (EDA)\end{tabular} & \begin{tabular}[c]{@{}c@{}}FN\%\\ (RDA)\end{tabular} & \begin{tabular}[c]{@{}c@{}}FP\%\end{tabular} & \begin{tabular}[c]{@{}c@{}}FN\%\\ (EDA)\end{tabular} & \begin{tabular}[c]{@{}c@{}}FN\%\\ (RDA)\end{tabular} \\ \hline
30 & 6.95 & 1.69 & 2.53 & 4.22 & 0.97 & 3.94 & 0.19 & 0.00 & 0.95 \\ \hline
60 & 7.59 & 2.73 & 3.05 & 5.77 & 2.27 & 2.73 & 0.1 & 0.00 & 0.46 \\ \hline
90 & 6.99 & 2.55 & 2.86 & 5.27 & 2.17 & 2.3 & 0.06 & 0.00 & 0.08 \\ \hline
120 & 7.27 & 2.24 & 2.45 & 4.15 & 1.77 & 2.05 & 0.05 & 0.00 & 0.01 \\ \hline
150 & 6.81 & 2.16 & 2.36 & 4.03 & 1.93 & 2.08 & 0.00 & 0.00 & 0.00 \\ \hline
20000 & 0.89 & 0.41 & 0.43 & 0.00 & 0.00 & 0.00 & 0.00 & 0.00 & 0.00 \\ \hline
40000 & 0.11 & 0.00 & 0.00 & 0.0 & 0.00 & 0.00 & 0.0 & 0.00 & 0.00 \\ \hline
\end{tabular}
\caption{Table showing the false positive and the false negative $\%$ for different $n$ for simple, $\Delta$, and filtered-$\Delta$-mean-difference algorithms done over 10,000 trials for the smartgrid dataset with $\epsilon = 40$.  For filtration $\Delta_{th} = 200$.}
\label{tab:fp_house}
\vspace{-4mm}
\end{center}
\end{table*}

\section{Experiments and Evaluations}

In this section, we run our proposed detection algorithms on some real-world datasets and evaluate their performance.  %
In each case, we observe the sensor readings for a duration that it takes to obtain $n$ samples, after which the algorithm outputs the presence/absence of an attacker.

\para{Simulated Attacks} To test the effectiveness of our proposed detection algorithm, we consider two types of attacks. In the first attack, the attacker predicts the exact sensor readings (without the addition of $\epsilon$) and uses that as the faked sensor readings to impersonate the compromised sensor. We refer to this attack as the ``Exact Duplication Attack" (EDA). In the second form of attack, to the exact prediction the attacker randomly injects micro-distortions, i.e., adds or subtracts $\epsilon$ randomly. We refer to this attack as the ``Random Distortion Attack" (RDA).

\begin{figure}[htb!]
\centering
\includegraphics[width=0.7\linewidth]{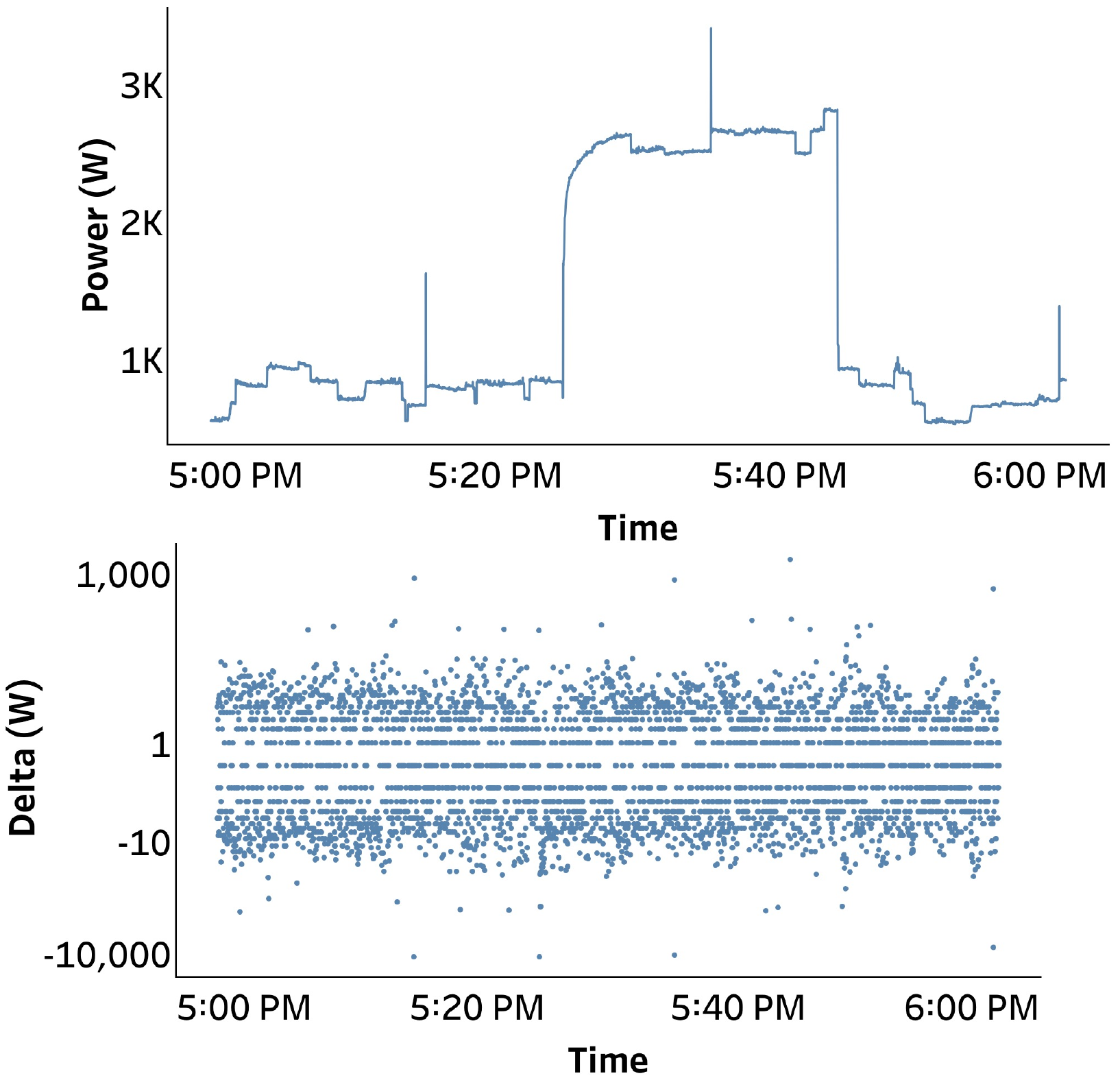}

\caption{ Figure depicting typical fluctuations in smart grid dataset. The variation in $\Delta$ is plotted in $\log$ scale.}
\label{fig:house}
\end{figure}

\addtolength{\topmargin}{0.05in}
\subsection{Attack Detection in Smart Grid Meter Dataset} \label{sec:house}
We make use of the publicly available ``Rainforest Automation Energy Dataset for Smart Grid Meter Data" \cite{DVN/ZJW4LC_2017} that contains 1 Hz data %
from residential households.
We consider the total power usage data for House 1 (with 72 days of data) given by its sensor `mains'. %
The data statistics and the statistics of the associated $\Delta$-sequence are given in Table \ref{tab:house_data}. Also, see Figure \ref{fig:house} for typical fluctuations in the power consumption of the house and the generated $\Delta$-sequence.

Based on the data,  we choose $\epsilon$ to be 40 W ($\approx0.5\%$ of the maximum power usage). From our experiments (see Table \ref{tab:fp_house}), we observe that even for a very small sample size of 30,  the filtered-$\Delta$-mean-difference algorithm gives us a very good FP/FN $\%$ of less than $1\%$,  i.e., with around 99.9 percent accuracy,  the algorithm can detect the presence of an attacker (if any) in less than 30 seconds.  %
This case also highlights the advantage of the filtration step that essentially removes the sudden changes in the power drawn when some appliance is turned on (or off).  We see that without such a filtration step, the FP/FN $\%$ is around $4.2\%$ for a similar duration of 30 seconds. Notice from Table \ref{tab:house_data}, that the maximum $\Delta$ is quite large. Filtering these high $\Delta$ values reduces the variance of the considered $\Delta'$ sequence allowing for faster and accurate detection. %
In contrast, the simple mean difference algorithm required around 20,000 samples to achieve a similar level of accuracy of less than $1\%$.

\begin{table}[]
\begin{center}
\begin{tabular}{|c|c|c|}
\hline
& Solar Power (kW) & $\Delta$-sequence (kW)\\ \hline
Maximum Value & 1576.54  &  1132.67\\ \hline
Minimum Value &  0.0  &  -925.18\\ \hline
Average Value & 394.118  &  $9.30 \times 10^{-5}$\\ \hline
Median Value &  276.36  &  -0.519\\ \hline
\end{tabular}
\end{center}
\caption{Table showing the power output data statistics for the solar plant and its associated $\Delta$-sequence.}
\label{tab:solar_data}
\end{table}

\subsection{Attack Detection in Solar Power Dataset}

\begin{figure}[h!]
  \centering
    \includegraphics[width=0.7\linewidth]{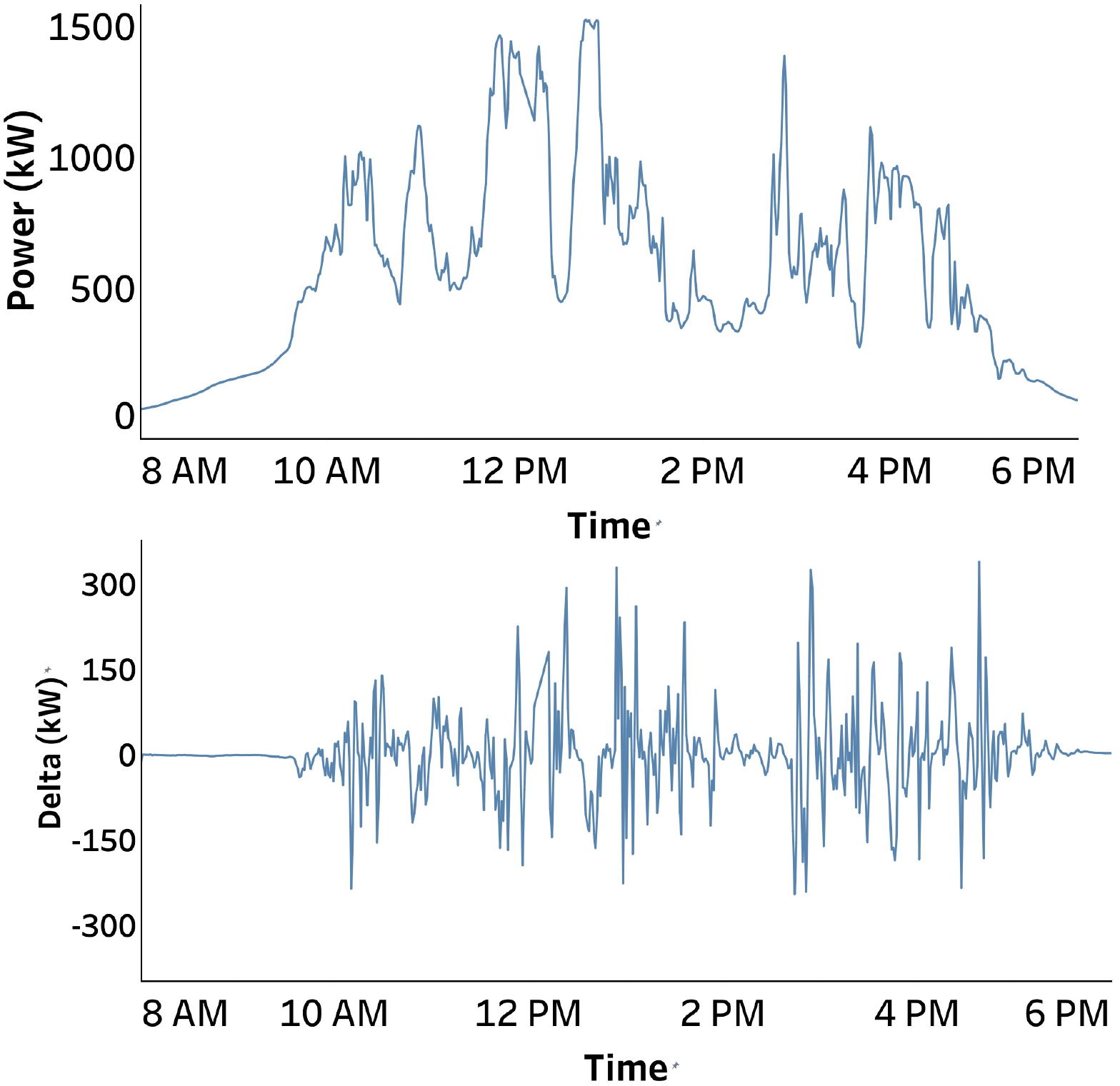}
  \caption{The solar power output and $\Delta_i$ fluctuation over a day.}
  \label{fig:solar}
\end{figure}

The solar power or the PV dataset is collected from a solar plant deployment in Singapore.  The data contains minute-wise values of the power generated from 7 stations from a period of 1/05/2020 to 17/06/2020, with the power generated given in kW for each station and the aggregate power output of the solar plant. 
We consider the sensor giving the aggregate power output of the solar plant to run our experiments.
The data statistics and the statistics of the associated $\Delta$-sequence are given in Table \ref{tab:solar_data}. Also, see Figure \ref{fig:solar} to see typical fluctuations in the solar data and the generated $\Delta$-sequence.

\begin{table*}[h]

\begin{center}
\begin{tabular}{|c|c|c|c|c|c|c|c|c|c|}
\hline
\multirow{2}{*}{$n$} & \multicolumn{3}{c|}{\begin{tabular}[c]{@{}c@{}}Simple  Mean Difference\end{tabular}} & \multicolumn{3}{c|}{\begin{tabular}[c]{@{}c@{}}$\Delta$-Mean-Difference\end{tabular}} & \multicolumn{3}{c|}{\begin{tabular}[c]{@{}c@{}}Filtered-$\Delta$-Mean-Difference\end{tabular}} \\ \cline{2-10} 
 & \begin{tabular}[c]{@{}c@{}}FP \%\end{tabular} & \begin{tabular}[c]{@{}c@{}}FN \%\\ (EDA)\end{tabular} & \begin{tabular}[c]{@{}c@{}}FN \%\\ (RDA)\end{tabular} & \begin{tabular}[c]{@{}c@{}}FP \%\end{tabular} & \begin{tabular}[c]{@{}c@{}}FN \%\\ (EDA)\end{tabular} & \begin{tabular}[c]{@{}c@{}}FN \%\\ (RDA)\end{tabular} & \begin{tabular}[c]{@{}c@{}}FP \%\end{tabular} & \begin{tabular}[c]{@{}c@{}}FN \%\\ (EDA)\end{tabular} & \begin{tabular}[c]{@{}c@{}}FN \%\\ (RDA)\end{tabular} \\ \hline
30 & 69.82 & 12.3 & 12.99 & 0.0 & 0.0 & 1.79 & 0.0 & 0.0 & 1.29 \\ \hline
60 & 71.76 & 13.39 & 13.59 & 0.0 & 0.0 & 0.20 & 0.0 & 0.0 & 0.12 \\ \hline
90 & 74.15 & 13.78 & 13.84 & 0.0 & 0.0 & 0.01 & 0.0 & 0.0 & 0.01 \\ \hline
120 & 75.83 & 14.74 & 14.75 & 0.0 & 0.0 & 0.0 & 0.0 & 0.0 & 0.0 \\ \hline
600 & 76.43 & 18.36 & 18.47 & 0.0 & 0.0 & 0.0 & 0.0 & 0.0 & 0.0 \\ \hline
\end{tabular}
\caption{Table showing the false positive and the false negative $\%$ for different $n$ for simple, $\Delta$ and filtered-$\Delta$-mean-difference algorithms done over 10,000 trials for the Solar Dataset with $\epsilon = 7.5$.  For filtration $\Delta_{th} = 30$.}
\vspace{-5mm}
\label{tab:fp_solar}
\end{center}
\end{table*}

 We know that at night time the solar power output sensor readings are all zeros, where the presence of an attacker can be detected extremely fast. Hence, our evaluation only considers the more challenging daytime values (from $8$am to $6$pm), which is 10 hours, or 600 points (per-minute) with a chosen $\epsilon$ as $7.5$kW (or 0.5\% of the maximum output of around $1.5$MW in this solar plant). As such, for each algorithm, we evaluate at most 600 samples after which we assume that the attacker can be detected. Even at daytime, we observe from Table \ref{tab:fp_solar} that we can detect an attacker with relatively high accuracy (of around 99.8$\%$) with 60 samples (that translates to 1hr time) for the $\Delta$ mean based algorithms both with and without the filtration step. In this case, %
 we see that though the filtration step helps, it does not have a significant impact on the accuracy as we saw earlier in Section \ref{sec:house}.  In addition, we also see that the simple mean difference algorithm for these short intervals performs even worse than random guessing. 
 
\begin{comment}
One way of improving detection accuracy that we show in the previous section was to increase the number of samples. Another way would be to increase the amount of induced distortion. In general, a higher distortion would help in obtaining a significant statistical difference faster.
%
We observe that, to obtain a FP/FN $\%$ of $1\%$ with only 600 samples, an $\epsilon$ of $75$ (10 times the size of the micro-distortion) is needed, and for FP/FN $\%$ of $0.1\%$ the needed $\epsilon$ is $120$.

\end{comment} 

%

%
%
%
%
%
%
%
%
%
%
%
%
%
%
%
%
%

%
%
%
%
%
%
%
%

%
%
%
%
%
%
%
%
%
%
%
%
%
%
%
%
%
%
%
%
%
%
%
%
%
%
%
%
%

%
\addtolength{\topmargin}{0.04in}
\section{Conclusion}
In this paper, we present a micro-distortion based detection algorithm that can help in a fast and accurate detection of stealthy attackers, with  low-cost changes to legacy systems.
A future goal is to develop other schemes that also work for rapidly fluctuating volatile systems. 
In conjunction with the current work, such a solution can be made to work in a wide range of scenarios to prevent stealthy attacks.

\section{Acknowledgements}
This research is supported in part by the National Research Foundation, Prime Minister's Office, Singapore under the Energy Programme and administrated by the Energy Market Authority (EP Award No. NRF2017EWT-EP003-047) 
and under its Campus for Research Excellence and Technological Enterprise (CREATE) programme, and in part by SUTD Start-up Research Grant (SRG Award No: SRG ISTD 2020 157).

\bibliographystyle{IEEEtran}
\begin{small}
\bibliography{bibliography}
\end{small}

\end{document}